\documentclass[conference,letterpaper]{IEEEtran}
\usepackage{graphicx,amsmath,latexsym,amssymb}
\ifCLASSINFOpdf
\else
\fi
\begin{document}

\title{Doubly Massive mmWave MIMO Systems: Using Very Large Antenna Arrays at Both Transmitter and Receiver}

\author{\IEEEauthorblockN{Stefano Buzzi  and Carmen D'Andrea}
\IEEEauthorblockA{DIEI - Universit\`a di Cassino  e del Lazio Meridionale\\
I-03043 Cassino (FR) - Italy \\
{\tt \{buzzi, carmen.dandrea\}@unicas.it}}
}


\maketitle

\begin{abstract}
One of the key features of next generation wireless communication systems will be the use of frequencies in the range 10-100GHz (aka mmWave band) in densely populated indoor and outdoor scenarios.  Due to the reduced wavelength, antenna arrays with a large number of antennas can be packed in very small volumes, making thus it possible to consider, at least in principle, communication links wherein not only the base-station, but also the user device, are equipped with very large antenna arrays. We denote this configuration as a ``doubly-massive'' MIMO wireless link. This paper introduces the concept of doubly massive MIMO systems at mmWave, showing that at mmWave the fundamentals of the massive MIMO regime are completely different from what happens at conventional sub-6 GHz cellular frequencies. 
It is shown for instance that  the multiplexing capabilities of the channel and its rank are no longer ruled by the number of transmit and receive antennas, but rather by the number of scattering clusters in the surrounding environment. 
The implications of the doubly massive MIMO regime on the transceiver processing, on the system energy efficiency and on the system throughput are also discussed.
\end{abstract}


%
\IEEEpeerreviewmaketitle

\section{Introduction}
The use of frequency bands in the range $10-100$GHz, a.k.a. millimeter wave (mmWave), for cellular communications, is among the most striking technological innovations brought by  fifth generation (5G)  wireless networks \cite{Buzzi5G}. Classically, mmWave carrier frequencies have been neglected for cellular communications in favour of sub-6 GHz bands, due to both their  path-loss, that increases quadratically with the carrier frequency, and to their increased atmospheric absorption. The scarcity of available frequency bands in the sub-6 GHz spectrum has been the main thrust for reconsidering this conventional wisdom and indeed recent research \cite{rappaport2013millimeter,HeathBook} has shown that mmWave can be actually used for cellular communications mainly for two reasons:
(a) the advent of heterogeneous networks and the progressive miniaturization of the radio cells  has brought increased attention on short-range (up to 100 - 200 meters) communications: for such distances, the increased atmospheric absorption due to humidity, rain and fog for mmWave frequencies has a negligible impact, tipically less than 1dB; and, (b) it has been recognized that, although the path-loss increases proportionally to $1/\lambda^2$, for a given physical antenna aperture, the maximum directional gains are proportional to $1/\lambda^2$ as well, thus implying that the scaling in the transmit and receive antenna gains overcomes the path-loss increase.
The latter argument leads to the conclusion that mmWave transceivers must be necessarily equipped with multiple antennas, so that MIMO processing is one distinguishing and inescapable feature of mmWave systems. 

For conventional sub-6 GHz cellular systems it has been shown that equipping a base station (BS) with a very large ($>100$) number of antennas, a technique usually referred to as \textit{Massive MIMO} \cite{Marzetta10,SILarssonMM}, brings a huge increase in the network capacity, mainly due to  the capability of serving several users on the same frequency slot with nearly orthogonal vector channels. In the massive MIMO literature, while the number of antennas at the BS grows large, the user device is usually assumed to 
have only one or very few antennas: indeed, at sub-6 GHz frequencies the wavelength is in the order of several centimeters, thus making it difficult to pack many antennas on small-sized user devices. When moving to mmWave, however, the wavelength gets reduced, and, at least in principle, a large number of antennas can be mounted not only on the BS, but also on the user device. As an example, at a carrier frequency of 30GHz the wavelength is 1 cm, 
and for a planar antenna array with $\lambda/2$ spacing, more than 180 antennas can be placed in an area as large as a standard credit card; this number climbs up to 1300 at a carrier frequency of 80 GHz. This leads to the concept of \textit{doubly massive MIMO system} \cite{buzziWCNC2016keynote}, that is defined as a wireless communication system where the number of antennas grows large at both the transmitter and  the receiver. 
While being aware that there are a number of serious practical constraints -- e.g.,  large power consumption, low efficiency of power amplifiers, hardware complexity, ADC and beamformer implementation -- that currently prevent the feasibility of an user terminal equipped with a very large number of 
antennas\footnote{Note however that the doubly massive MIMO scenario is of interest also for modeling wireless fronthaul networks wherein both the anchor BS and the remote BSs may be equipped with large antenna arrays.}, 
we believe that these are just technological issues that will be solved or worked around in the near future, and this paper presents  results on the doubly massive MIMO regime for wireless systems operating at mmWave.  The contributions of this paper can be summarized as follows:
\begin{itemize}
\item[-]
We analyze the implications that the mmWave clustered channel model has on the multiplexing capabilities of the channel in the doubly massive MIMO regime. In particular, we show that, differently from what usually happens at sub-6 GHz frequencies, increasing the number of transmit and receive antennas yields a limited multiplexing gain, which instead is mainly tied to the number of reflecting clusters. We show instead that the received signal power scales linearly with the product of the number of transmit and receive antennas.
\item[-]
The energy efficiency of a wireless system operating in the doubly massive MIMO regime is also studied. It is shown that the operating point  maximizing the system energy efficiency actually depends on the amount of energy consumed by the transmitter circuits, thus confirming  the relevance of hardware issues to the real success of mmWave wireless communications.
\item[-]
Initial considerations on the implications of the clustered channel model with regard to issues such as the optimality of analog beamforming, the complexity of channel estimation, and the multiplexing capabilities of the system are also discussed for the doubly massive MIMO mmWave regime. 
\end{itemize}

This paper is organized as follows. In the next Section the considered system model is described, including the clustered channel model. Section III contains the analysis of the doubly massive MIMO regime, from both the points of view of spectral and energy efficiency. Section IV is devoted to the discussion of the numerical results, while, finally, in Section V concluding remarks are given, along with a discussion on the further work that is needed in order to gain a better understanding of the doubly massive MIMO regime at mmWave.

\section{The system model}

We consider the downlink of a single-cell multiuser MIMO (MU-MIMO) system wherein one BS communicates, on the same frequency slot, with several mobile users. 
We denote by $N_T$ the number of transmit antennas at the BS, and by $N_R$ the number of receive antennas at the user's device\footnote{For the sake of simplicity we assume that all the mobile receivers have the same number of antennas; however, this hypothesis can be easily relaxed.}. 

\subsection{The clustered channel model}

We focus on a narrowband clustered channel model\footnote{References \cite{spatiallysparse_heath,cairewsa2016,buzziwsa2016,bogale2014beamforming,liang2014low,lee2014exploiting} are a sample of recent papers - by different set of authors - that have embraced the clustered channel model for modelling MIMO propagation at mmWave carrier frequencies.}, i.e. either the multipath delay spread is much smaller than the inverse of the signal bandwidth, or the channel has been made frequency flat through OFDM modulation and FFT processing.  In both cases, the baseband equivalent of the propagation channel between the transmitter and the generic receiver\footnote{For ease of notation, we omit, for the moment, the subscript $"k"$ to denote the BS to the $k$-th user channel matrix.}
is  represented by an $(N_R \times N_T)$-dimensional matrix expressed as:
\begin{equation}
\mathbf{H}=\gamma\sum_{i=1}^{N_{\rm cl}}\sum_{l=1}^{N_{{\rm ray},i}}\alpha_{i,l}
\sqrt{L(r_{i,l})} \mathbf{a}_r(\phi_{i,l}^r) \mathbf{a}_t^H(\phi_{i,l}^t) + \mathbf{H}_{\rm LOS}\; .
\label{eq:channel1}
\end{equation}
In Eq. \eqref{eq:channel1}, 
we are implicitly assuming that the propagation environment is made of $N_{\rm cl}$ scattering clusters, each of which contributes with $N_{{\rm ray}, i}$ propagation paths, $i=1, \ldots, N_{cl}$, plus a  possibly present LOS component.  
We denote by  $\phi_{i,l}^r$ and $\phi_{i,l}^t$ the angles of arrival and departure of the $l^{th}$ ray in the $i^{th}$ scattering cluster, respectively. 
The quantities $\alpha_{i,l}$ and $L(r_{i,l})$ are the complex path gain and the attenuation associated  to the $(i,l)$-th propagation path. 
The complex gain  $\alpha_{i,l}\thicksim \mathcal{CN}(0, \sigma_{\alpha,i}^2)$, with  $\sigma_{\alpha,i}^2=1$  \cite{spatiallysparse_heath}. The factors $\mathbf{a}_r(\phi_{i,l}^r)$ and $\mathbf{a}_t(\phi_{i,l}^t)$ represent the normalized receive and transmit array response vectors evaluated at the corresponding angles of arrival and departure; for an uniform linear array (ULA) with half-wavelength inter-element spacing we have:
\begin{equation}
\mathbf{a}_t(\phi_{i,l}^t)=\displaystyle \frac{1}{\sqrt{N_T}}[1 \; e^{-j\pi \sin \phi_{i,l}^t} \; \ldots \; e^{-j\pi (N_T-1) \sin \phi_{i,l}^t}] \; ;
\label{eq:ULA}
\end{equation}
A similar expression can be also given for $\mathbf{a}_r(\phi_{i,l}^r)$.
Finally, $\gamma=\displaystyle\sqrt{\frac{N_R N_T}{\sum_{i=1}^{N_{\rm cl}}N_{{\rm ray},i}}}$  is a normalization factor ensuring that the received signal power scales linearly with the product $N_R N_T$.
Regarding the LOS component, denoting by 
$\phi_{\rm LOS}^r$,  $\phi_{\rm LOS}^t$,
 the arrival and departure angles corresponding to the LOS link, we assume that
\begin{equation}
\begin{array}{llll}
\mathbf{H}_{\rm LOS} = &  
I_{\rm LOS}(d) \sqrt{N_R N_T L(d)} e^{j \eta} \mathbf{a}_r(\phi_{\rm LOS}^r)  \mathbf{a}_t^H(\phi_{\rm LOS}^t) \; .
\end{array}
\label{eq:Hlos}
\end{equation}
In the above equation, $\eta \thicksim \mathcal{U}(0 ,2 \pi)$, while $I_{\rm LOS}(d) $ is a random variate indicating if a LOS link exists between transmitter and receiver, with $p$ the probability that $I_{\rm LOS}(d) =1$.
A detailed description of all the parameters needed for the generation of sample realizations for the channel model of Eq. \eqref{eq:channel1} is reported in \cite{buzzidandreachannel_model}, and we refer the reader to this reference for further details on the  channel model. 

\begin{figure}[!t]
\centering
\includegraphics[scale=0.22]{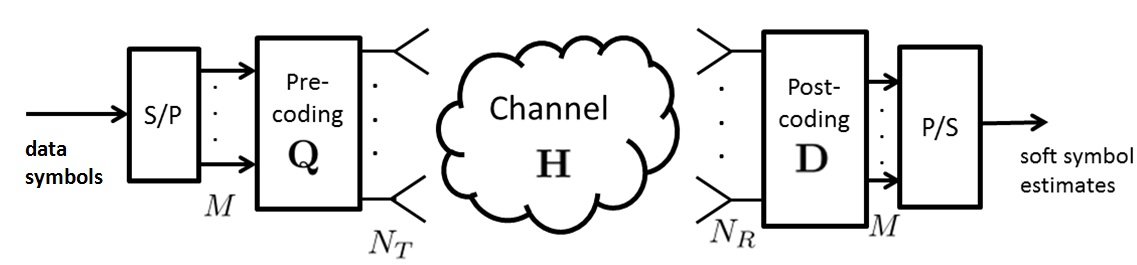}
\caption{The considered transceiver model in the narrowband case.}
\label{Fig:scheme}
\end{figure}

\subsection{Transmitter and Receiver Processing}
We consider a multiuser model representing the downlink of a cellular system (the single-user scenario is
depicted in Fig. \ref{Fig:scheme}). Let $K$ be the number of users simultaneously served by the BS, and denote by $M$ the number of data symbols sent to each user in each signalling interval\footnote{Otherwise stated, the BS transmits in each time-frequency slot $M K$ data symbols.}.  Denoting by $\mathbf{x}_k$ the $M$-dimensional vector of the data symbols intended for the $k$-th user, the discrete-time signal transmitted by the BS is expressed as the $N_T$-dimensional vector
$\mathbf{s}_T=  \sum_{k=1}^{K} \mathbf{Q}_k \mathbf{x}_k$, with $\mathbf{Q}_k$ the $(N_T \times M)$-dimensional precoding matrix for the $k$-th user. The signal received by the generic $k$-th user is expressed as the following $N_R$-dimensional vector
\begin{equation}
\mathbf{y}_k= \mathbf{H}_k \mathbf{s}_T + \mathbf{w}_k \; ,
\label{eq:yk}
\end{equation}
with $\mathbf{H}_k$ representing the channel from the BS to the $k$-th user and $\mathbf{w}_k$ is the $N_R$-dimensional additive white Gaussian noise with zero-mean i.i.d. entries with variance $\sigma_n^2$. Denoting by $\mathbf{D}_k$ the $(N_R \times M)$-dimensional post-coding matrix at the $k$-th user device, the following $M$-dimensional vector is finally obtained:
\begin{equation}
\mathbf{r}_k= \mathbf{D}_k^H \mathbf{H}_k \mathbf{Q}_k \mathbf{x}_k + \sum_{\ell =1, \ell \neq k}^K
\mathbf{D}_k^H \mathbf{H}_k \mathbf{Q}_\ell \mathbf{x}_\ell + \mathbf{D}_k^H \mathbf{w}_k  \; .
\end{equation}

\subsubsection{Channel matched hybrid (CM-HY) beamforming}
Letting $\mathbf{H}_k= \mathbf{U}_k \mathbf{\Lambda}_k \mathbf{V}_k^H$ denote the singular-value-decomposition (SVD) of the matrix $\mathbf{H}_k$, the $k$-th user precoding and post-coding matrices $\mathbf{Q}_k$ and  $\mathbf{D}_k$  could be chosen\footnote{For the sake of simplicity we are not considering here more sophisticated forms of beamforming such as the zero-forcing strategy.} as the columns of the matrices $\mathbf{V}_k$ and 
$\mathbf{U}_k$, respectively, corresponding to the largest entries in the  eigenvalue matrix $\mathbf{\Lambda}_k$. 
This, however, would require a fully digital structure with a number of RF chains equal to the number of antennas; in order to lower the system complexity, we resort to a CM-HY structure, wherein the number of RF chains in the BS is equal to $KM$, while at the mobile terminal it is equal to $M$, and, in order to approximate the ideally desired precoding and post-coding matrices, the  block coordinate descent for subspace decomposition algorithm is used \cite{ghauch2015subspace}. 

\subsubsection{Pure analog (AN) beam-steering beamforming}
We also consider beam-steering analog beamforming, i.e.  we assume that the columns of the precoding and post-coding matrices are unit-norm beam-steering vectors expressed as in Eq. \eqref{eq:ULA}. The columns of the precoding matrix  are chosen as the array responses corresponding to the departure angles in the channel model \eqref{eq:channel1} associated to the $M$ dominant paths, and, similarly, the columns of the post-coding matrix are the array responses corresponding to the $M$ arrival angles associated to the $M$ dominant paths. In order to avoid self-interference, we have included a further constraint in the choice of the dominant paths to ensure that the angles of departure (arrivals) of the selected paths are spaced of at least 5 deg.

\subsection{The considered performance measures}
We will consider two performance measures: the achievable spectral efficiency (ASE) and the achievable spectral energy efficiency (ASEE). 
The ASE is measured in [bit/s/Hz], while the ASEE is measured in 
[bit/Joule/Hz]; multiplying the ASEE by the communication bandwidth we obtain the usual energy efficiency performance metric measured in [bit/Joule] \cite{buzziJSAC2016}.
 Assuming Gaussian data symbols\footnote{The impact on the ASE of a finite-cardinality modulation is a topic worth future investigations.}, it can be shown that the ASE can be expressed as
\begin{equation}
{\rm ASE}= \displaystyle \sum_{k=1}^K \log \det \left[ \mathbf{I}_M + \frac{P_T}{KM}\mathbf{R}_{\overline{k}}^{-1}\mathbf{D}_k^H \mathbf{H}_k
\mathbf{Q}_k \mathbf{Q}_k^H \mathbf{H}_k^H \mathbf{D}_k
\right] \; ,
\end{equation}
 wherein $\mathbf{I}_M$ is the identity matrix of order $M$, $P_T$ is the BS transmit power, and 
 $\mathbf{R}_{\overline{k}}$ is the covariance matrix of the overall disturbance seen by the $k$-th user receiver, i.e.:
\begin{equation}
\mathbf{R}_{\overline{k}}=\sigma^2_n \mathbf{D}_k^H \mathbf{D}_k +
\frac{P_T}{MK} \sum_{\ell =1, \ell \neq k}^K
\mathbf{D}_k^H \mathbf{H}_k \mathbf{Q}_\ell \mathbf{Q}_\ell^H \mathbf{H}_k^H \mathbf{D}_k \; .
\end{equation}
Regarding the ASEE, denoting by $P_c$ the amount of power consumed by the transmitting circuitry for each transmit antenna, we have the following definition:
\begin{equation}
{\rm ASEE}= \displaystyle \frac{\rm ASE}{N_T P_c + \eta P_T} \; ,
\label{eq:ASEE}
\end{equation}
with $\eta>1$ a scalar coefficient modelling the power amplifier inefficiency.

\section{The doubly massive MIMO regime}
In this section we focus on the doubly massive regime, namely we assume that both $N_T$ and $N_R$ grow large and discuss the implications of the adopted clustered channel model. The main goal of this section is to highlight that MIMO propagation at mmWave, as reflected from the clustered channel model,  is fundamentally different from the propagation mechanisms at sub-6 GHz frequencies, so that the analysis of the (doubly) massive MIMO regime at mmWave is not a straightforward extension of the massive MIMO regime at conventional sub-6 GHz cellular frequencies\footnote{The discussion will be concise due to lack of space, and extensive analytical results will be provided elsewhere.}.

\subsection{The channel rank}
Given the channel model in Eq. \eqref{eq:channel1}, it is seen that, including the LOS component \eqref{eq:Hlos}, 
and assuming for simplicity that $N_{{\rm ray},i}=N_{\rm ray} \; \forall i=1 \ldots N_{\rm cl}$,
the channel has at most rank $N_{\rm cl} N_{\rm ray} +1$ since it is expressed as the sum of $N_{\rm cl} N_{\rm ray} +1$ rank-1 matrices. Otherwise stated, while in sub-6 GHz the rich scattering environment assumption leads to modelling the channel matrix $\mathbf{H}$ as having i.i.d. entries contributing to a rank that is equal to $\min(N_T,N_R)$ with probability 1, in the case of mmWave
the channel rank depends on the propagation scene geometry, and is independent of the number of antennas (we will see in a moment  this statement partly confuted). 
The multiplexing capabilities of the channel, thus, do not grow linearly with the minimum between $N_T$ and $N_R$, but rather depend on the number of clusters and scatterers in the propagation environment. The number of antennas just contribute to the increase of the received power, which indeed can be shown to increase proportionally to the product $N_TN_R$.

That said, we can conjecture that, for increasing number of antennas, the directive beams become narrower and narrower, and more scatterers can be resolved, thus implying that channel rank increases (even though probably not linearly) with the number of antennas. We highlight however that ours is a conjecture that would need experimental validation. 

\subsection{The channel eigenmodes}
Given the \textit{continuous} random location of the scatterers, the set of arrival angles will be different with probability 1, i.e. there is a zero probability that two distinct scatterers will contribute to the channel with the same departure and arrival angles. Since, 
for large number of antennas, we have that 
 \begin{equation}
 \mathbf{a}_x(\phi_p^x)^H  \mathbf{a}_x(\phi_q^x) \rightarrow 0 \; ,
  \end{equation}
provided that  $(\phi_p^x) \neq (\phi_q^x)$, we can conclude that 
for large $N_T$, the vectors $\mathbf{a}_t(\phi_{i,l}^t)$, for all $i$ and $l$, converge to an orthogonal set; similarly, the vectors $\mathbf{a}_r(\phi_{i,l}^r)$, for all $i$ and $l$, converge to an orthogonal set as well. 
Accordingly, in the doubly massive MIMO regime, the array response vectors 
$\mathbf{a}_r(\cdot)$ and $\mathbf{a}_t(\cdot)$ 
become the left and right singular vectors of the channel matrix, i.e. the channel representation in \eqref{eq:channel1} coincides with its SVD. Under this situation, we have three key results:
(a) The optimum precoding and post-coding matrices coincide with the AN beamforming structures outlined in the previous section;
(b) the entries of the optimum precoding and post-coding matrices have all the same magnitude, i.e. they just differ in the phase, thus implying that analog pre-coding and post-coding exhibits no loss with respect to the ideal case of digital beamforming; and
(c) in a multiuser setting, the channels corresponding to different users are orthogonal (provided that there is a sufficient spacing between users and between the several angles of arrival and departure) and so interference-free multiuser transmission is in principle possible. 

\subsection{Leveraging sparse signal processing}
The channel model in \eqref{eq:channel1} is basically a parametric model; when doing channel estimation, thus, instead of separately estimating all the entries of the channel matrix, it is more convenient to estimate the channel parameters such as arrival and departure angles, and scattering coefficients. We have thus that the computational complexity of the channel estimation task does not grow in the product $N_R N_T$. 


\begin{figure}[!t]
\centering
\includegraphics[scale=0.30]{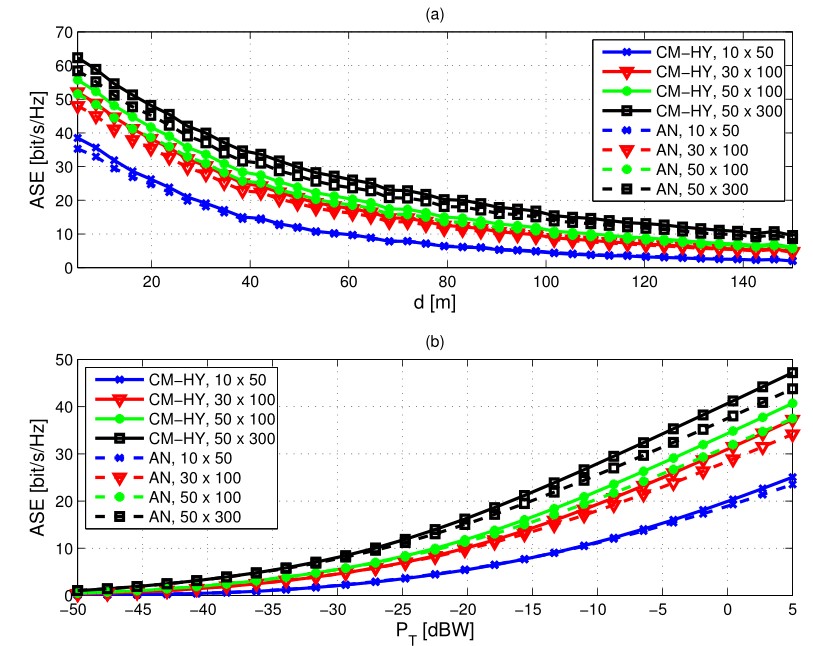}
\caption{(a) ASE versus link length for several values of $N_R \times N_T$, $P_T=0$dBW, $M=4$; (b) ASE versus transmit power for several values of $N_R \times N_T$, $d=30$m, $M=4$.}
\label{fig:Pd}
\end{figure}


\begin{figure}[!t]
\centering
\includegraphics[scale=0.30]{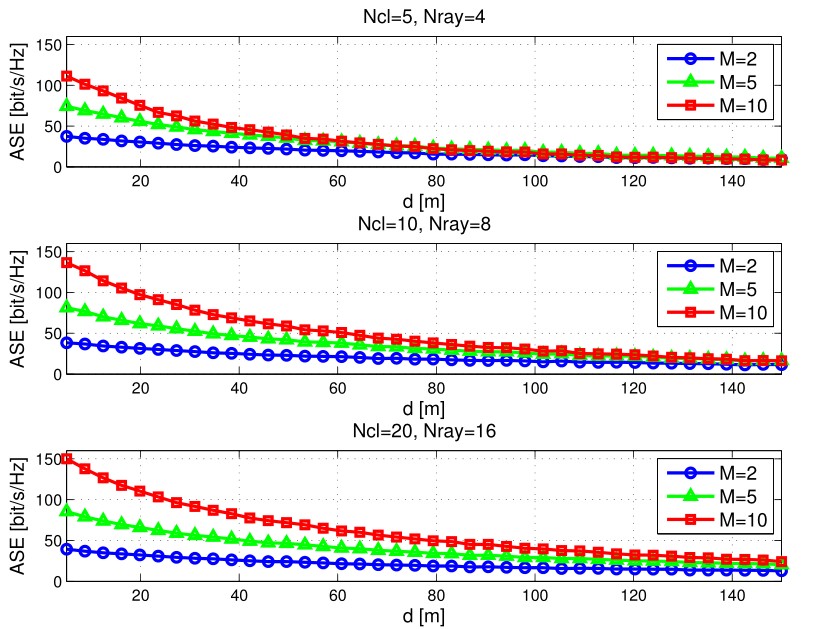}
\caption{Impact of the number of clusters. $P_T=0$dBW, $N_R \times N_T= 50 \times 100$.}
\label{fig:Ncl}
\end{figure} 

\section{Numerical results}
We now provide extensive simulation results showing the ASE and the ASEE for a wireless communication system operating at mmWave with a large number of antennas at both the transmitter and the receiver. For the sake of simplicity, we start by considering a single-user system. The parameters for the generation of the channel matrices are the ones reported in \cite{buzzidandreachannel_model} for the ``street canyon model''.
Fig. \ref{fig:Pd} reports the ASE versus the link length and the transmit power, respectively, for several values of $N_R \times N_T$, comparing the performances of Channel Matched Hybrid (CM-HY) and Analog (AN) Beamforming. Obviously, the ASE is an increasing function of the transmit power and a decreasing function of the link length; results confirm that the ASE improves for growing number of antennas, even though the performance improvements are not as striking as the ones observed at sub-6 GHz frequencies: as an example, from Fig. \ref{fig:Pd} (b) it is seen that for $P_T=5$ dBW, comparing the system with $N_R \times N_T= 50 \times 100$ with the system $50 \times 300$ (that is 3x more antennas), the ASE increases approximately by 20$\%$. As already discussed, this limited gain is due to the fact that the channel entries are no longer i.i.d. complex Gaussian variates, but they obey the parametric model \eqref{eq:channel1}.
In Section III we mentioned that the increase in the number of antennas essentially provides just a gain in the received signal power. This statement can be easily seen through a careful inspection of Fig. \ref{fig:Pd} (b); by shifting the several curves of an amount exactly equal to the theoretical gain granted by the increase in the number of antennas,  it is easy to see that all the curves approximately coincide, and this confirm that increasing the number of antennas does not provide additional degrees of freedom, but just provides a Signal-to-Noise Ratio gain\footnote{Remember however that this statement is no longer true of we assume that the number of reflecting clusters increases with the size of the antenna array.}.
 The fact that the multiplexing capabilities of the channel are ruled by the parameters $N_{\rm cl}$ and $N_{\rm ray}$ is confirmed by the plots in Fig. \ref{fig:Ncl}, wherein the ASE versus the link length for different values of $N_{\rm cl}$ and $N_{\rm ray}$ is reported; results clearly show that increasing the number of scatterers greatly improves the system ASE, especially for large values of the multiplexing order $M$; again, we remark that this is in sharp constrast with what happens at sub-6 GHz frequencies, wherein the multiplexing capabilities of the channel are tied to the number of transceiver antennas.
 
\begin{figure}[!t]
\centering
\includegraphics[scale=0.30]{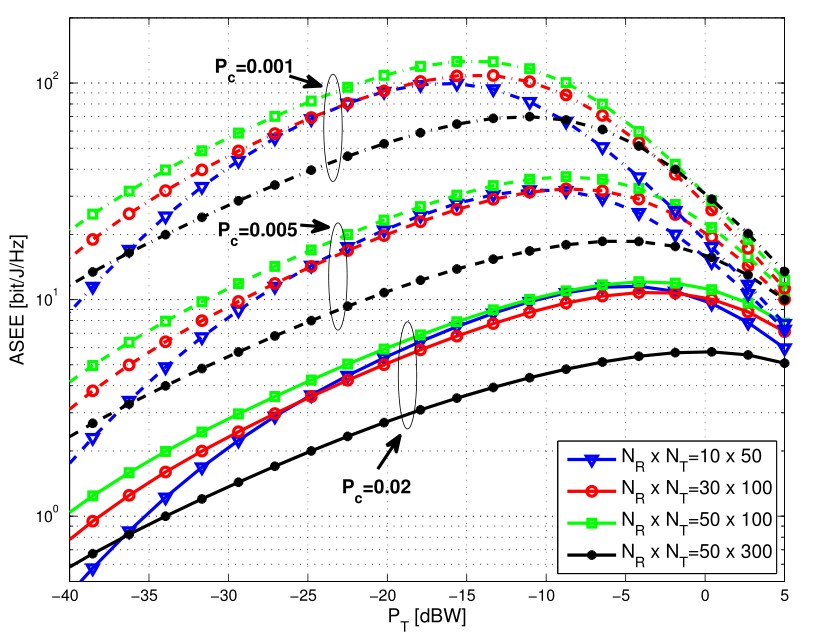}
\caption{Spectral EE versus transmit power for several values of $N_R \times N_T$ and for several values of $P_c$. $d=30$m, $M=4$.}
\label{fig:EE}
\end{figure}

\begin{figure}[!t]
\centering
\includegraphics[scale=0.30]{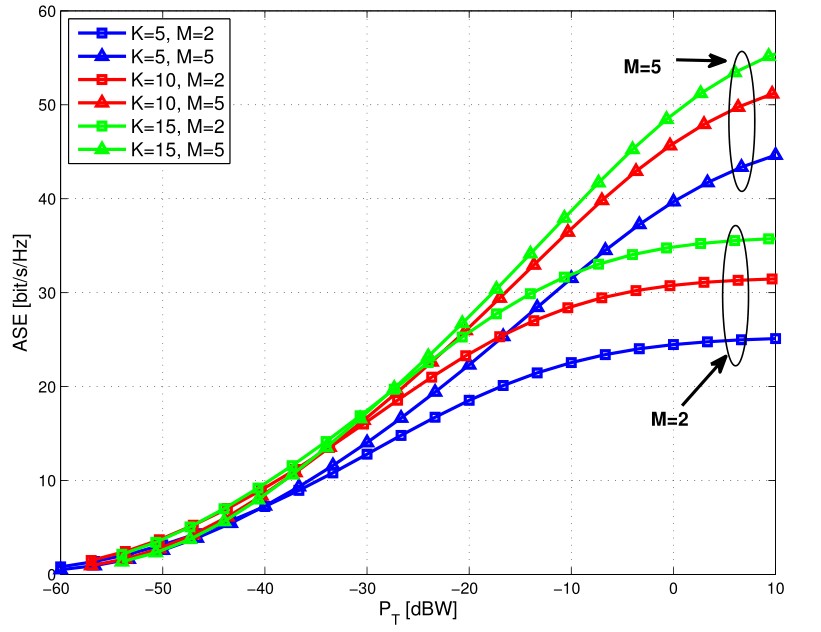}
\caption{ASE for a multiuser MIMO system for several numbers of users and for $M=2, 5$. $N_R \times N_T= 50 \times 200$. The users have been assumed to have a random location with a distance from the transmitter in the range $[10, 100]$ meters.}
\label{fig:MU_MIMO}
\end{figure}

Energy efficiency is considered in Fig. \ref{fig:EE}, wherein the ASEE versus the transmit power is reported for several values of $N_R \times N_T$ and for three different values of $P_c$ the amount of hardware consumed power for each transmit antenna. Two observations are in order here: (a) of course, the larger $P_c$, the smaller the ASEE; and (b) it is seen that the ASEE is not a monotonically increasing function of the number of antennas; this behaviour can be easily justified by noting that both the numerator and the denominator in Eq. \eqref{eq:ASEE} increase with $N_T$, and thus, given the operating scenario, there is an optimal number of antennas that maximizes the system energy efficiency. Finally, Fig. \ref{fig:MU_MIMO} reports the ASE for a multiuser system. Results confirm that the system is well capable of supporting several users in the same time-frequency slot, mainly due to the fact that, as explained in the previous section, for increasing number of antennas the transmit and receive array response vectors tend to become orthogonal. Further investigations are needed here in order to derive the maximum number of supported users as a function of the system parameters, as well as the 0.95-percentile rate for the generic mobile device.

\section{Concluding remarks and future work}
This paper has provided a preliminary study of the doubly massive MIMO regime for a cellular system operating at mmWave carrier frequencies. Starting from the widely accepted clustered channel model, we have shown that increasing the number of antennas brings performance improvements that are based on propagation mechanisms different from what happens at sub-6 GHz communications. We have seen for instance that the channel rank and multiplexing capabilities depend on the number of reflecting clusters rather than on the number of transmit and receive antennas, which instead have a direct effect on the scaling of the received power. The implications of the clustered channel model on the optimality of analog beamforming have also been discussed. 
Our results have confirmed that there is much to gain from the use of a large number of antennas at both sides of a wireless link operating at mmWave. Nonetheless, there is still much work to do in order to fully understand the potentialities and the challenges brought by the doubly massive MIMO regime. Possible research paths are briefly illustrated in the following.
\begin{itemize}
\item[-]
The results shown in this paper are based on the clustered channel model. Although, as already discussed, this is by now a widely accepted model that has been used worldwide by several independent research groups, measurements should be made with a large number of antennas at both the transmitter and the receiver to actually validate the hypothesis that the channel rank is only weakly related to the dimension of the arrays.
\item[-]
Regarding the system energy efficiency, we have discussed about the impact of the circuit consumed power $P_c$. To this end, accurate models describing hardware energy consumption should be provided in order to be able to perform a better characterization of massive MIMO mmWave systems. While some databases exist for sub-6 GHz RF chains (see, e.g., \cite{powermeter}), and some results on the energy consumption of mmWave systems begin to appear \cite{mendez2016hybrid}, additional work is needed in this area. 
\item[-]
Still about energy efficiency, our results  have shown that while increasing the number of antennas is always beneficial for the system ASE, the system energy efficiency does not monotonically improves with the number of antennas, mainly due to the power consumption from the circuitry, that we have assumed to scale \textit{linearly} with the number of antennas. Of course, results would have been more appealing is we could have had antenna arrays whose hardware consumed power scales sub-linearly with the number of transmit antennas. Is this possible? May this be achieved someday through the use of new materials and technologies? See the paper \cite{barousis2014massive} for some ideas on this topic.
\item[-]
We have only marginally tackled the issue of channel estimation. Although the parametric channel model simplifies the problem of channel estimation, in the sense that the number of parameters to be estimated does not scale with the product $N_T N_R$, design and analysis of efficient channel estimation schemes, taking advantage of the channel sparsity, is an interesting research path. 
\item[-]
Investigation on the use of antenna arrays with spacing less than $\lambda/2$ is also a topic worth being investigated: indeed, to save space on the user device, antennas might be packed more densely than the canonical $\lambda/2$ spacing. Intuition suggests that in this case a certain performance degradation should appear, but of course performance would still improve with increasing number of antennas. A characterization of such systems is still missing.
\end{itemize}

\bibliographystyle{IEEEtran}

\bibliography{FracProg_SB,finalRefs}

\end{document}